\documentclass[preprint2]{aastex}

\def\kms{\hbox{km s$^{-1}$}}
\def\VLSR{\hbox{$V_{\rm LSR}$}}
\def\hh{\hbox{$^{\rm h}$}}
\def\mm{\hbox{$^{\rm m}$}}
\def\ss{\hbox{$^{\rm s}$}}
\def\sun{\hbox{$\odot$}}

\def\lesssim{\mathrel{\hbox{\rlap{\hbox{\lower4pt\hbox{$\sim$}}}\hbox{$<$}}}}
\def\gtrsim{\mathrel{\hbox{\rlap{\hbox{\lower4pt\hbox{$\sim$}}}\hbox{$>$}}}}

\def\arcdeg{\hbox{$^\circ$}}
\def\arcmin{\hbox{$^\prime$}}
\def\arcsec{\hbox{$^{\prime\prime}$}}

\slugcomment{To appear in The Astrophysical Journal}

\shorttitle{Atomic Carbon and CO Isotope Emission}
\shortauthors{Oka et al.}

\begin{document}

\title{Atomic Carbon and CO Isotope Emission in the Vicinity of DR15}

\author{Tomoharu Oka\altaffilmark{1}, Satoshi Yamamoto\altaffilmark{1}, Mitsuhiro Iwata\altaffilmark{1}, 
Hiroyuki Maezawa\altaffilmark{1}, Masafumi Ikeda\altaffilmark{1}, Tetsuya Ito\altaffilmark{1}, 
Kazuhisa Kamegai\altaffilmark{1}, Takeshi Sakai\altaffilmark{1}, 
Yutaro Sekimoto\altaffilmark{2}, Ken'ichi Tatematsu\altaffilmark{2}, 
Yuji Arikawa\altaffilmark{3}, Yoshiyuki Aso\altaffilmark{2}, Takashi Noguchi\altaffilmark{2}, 
Sheng-Cai Shi\altaffilmark{2,4}, Keisuke Miyazawa\altaffilmark{2}, 
Shuji Saito\altaffilmark{5,6}, Hiroyuki Ozeki\altaffilmark{5,7}, Hideo Fujiwara\altaffilmark{5,8}, 
Masatoshi Ohishi\altaffilmark{9}, and Junji Inatani\altaffilmark{7}}

\altaffiltext{1}{{\footnotesize Research Center for the Early Univers and Department of Physics, The University of Tokyo, 7-3-1 Hongo, Bunkyo-ku, Tokyo 113-0033, Japan.}} 
\altaffiltext{2}{{\footnotesize Nobeyama Radio Observatory, National Astronomical Observatory of Japan, Nobeyama, Minamimaki, Minamisaku, Nagano 384-1305, Japan.}} 
\altaffiltext{3}{{\footnotesize Department of Astronomical Science, Graduate University for Advanced Studies, Nobeyama Radio Observatory, Nobeyama, Minamimaki, Minamisaku, Nagano 384-1305, Japan.}}
\altaffiltext{4}{{\footnotesize Purple Mountain Observatory, Nanjing, JiangSu 210008, China.}}
\altaffiltext{5}{{\footnotesize Institute for Molecular Science, Okazaki, Aichi 444-8585, Japan.}}
\altaffiltext{6}{{\footnotesize Fukui University, Fukui 910-8507, Japan.}} 
\altaffiltext{7}{{\footnotesize National Space Agency of Japan, Tsukuba, Ibaraki 305-8505, Japan.}} 
\altaffiltext{8}{{\footnotesize Department of Chemical System Engineering, University of Tokyo, Tokyo 113-0033 Japan.}}
\altaffiltext{9}{{\footnotesize National Astronomical Observatory of Japan, Mitaka, Tokyo 181-8588, Japan.}}

\email{tomo@taurus.phys.s.u-tokyo.ac.jp}

\begin{abstract}
We present observations of the $^3P_1$--$^3P_0$ fine structure transition of atomic carbon [CI], 
the {\it J}=3--2 transition of CO, 
as well as of the {\it J}=1--0 transitions of $^{13}$CO and C$^{18}$O 
toward DR15, an HII region associated with two mid-infrared dark clouds (IRDCs). 
The $^{13}$CO and C$^{18}$O {\it J}=1--0 emissions closely follow 
the dark patches seen in optical wavelength,
showing two self-gravitating molecular cores with masses of 2000 M$_{\sun}$ and 900 M$_{\sun}$, 
respectively, at the positions of the catalogued IRDCs.  

Our data show a rough spatial correlation between [CI] and $^{13}$CO {\it J}=1--0.     
Bright [CI] emission occurs in relatively cold gas behind the molecular cores, 
neither in highly excited gas traced by CO {\it J}=3--2 emission 
nor in HII region/molecular cloud interface.  
These results are inconsistent with those predicted by standard photodissociation region (PDR) models, 
suggesting an origin for interstellar atomic carbon unrelated to photodissociation processes.  
\end{abstract}


\section{INTRODUCTION}

The amount and distribution of atomic carbon is important for understanding of 
the physical and chemical structure, formation precesses, and thermal balance of 
interstellar molecular clouds. 
Since the ionization potential of neutral atomic carbon (11.266 eV) is close to 
the dissociation energy of CO (11.09 eV), 
it (C$^0$) is thought to be confined in a thin layer near the surfaces of molecular clouds.  
Chemical models of molecular clouds exposed to ultraviolet (UV) radiation 
demonstrate the concentration of C$^0$ in the range $A_{\rm V}\!=$3--6 mag 
(Tielens \&\ Hollenbach 1985; van Dishoeck \&\ Black 1988; 
Hollenbach, Takahashi, \&\ Tielens 1991).  

The $^3P_1$--$^3P_0$ fine-structure line of neutral atomic carbon [CI] 
(492.160651 GHz, Yamamoto \&\ Saito 1991) 
has been known to be a good tracer of interstellar atomic carbon.  
After the first detection (Phillips et al. 1980), 
the rapid advance in receiver sensitivity in the submillimeter wavelength has resulted in 
the $^3P_1$--$^3P_0$ line now being well studied in many Galactic sources.  
Large-scale observations have revealed that [CI] emission is widespread 
throughout molecular clouds (Plume, Jaffe, \&\ Keene 1994; Plume et al. 1999; 
Maezawa et al. 1999; Ikeda et al. 1999), 
being not confined in vicinities of ultraviolet sources.  
Observations on scales of tens of degrees has shown that 
[CI] emission is widespread also on galactic scales (Wright et al. 1991; Bennett et al. 1994).  

There have been some arguments (Keene et al. 1985; Schilke et al. 1993; Schilke et al. 1995; 
Maezawa et al. 1999; Ikeda et al. 1999) 
that C$^0$ exists also within deep inside molecular clouds 
where UV radiation is attenuated.  
Models of time-dependent chemistry 
(Suzuki 1979; Leung, Herbst, \&\ Huebner 1984; Suzuki et al. 1992; Lee et al. 1996) 
predict that C$^0$ can be abundant in the interior of a newly formed dense core, 
since the conversion timescale of C$^0$ to CO is comparable to the dynamical timescale of the core.  
Even in steady states, some models predict that, under certain circumstances, 
atomic carbon can be abundant in the absence of UV radiation 
(Pineau des For\^ets, Roueff, \&\ Flower 1992; 
Le Bourlot, Pineau des For\^ets, \&\ Roueff 1993; 
Le Bourlot, Pineau des For\^ets, Roueff, \&\ Flower 1995; 
Le Bourlot, Pineau des For\^ets, \&\ Roueff 1995; 
Lee, Roueff, Pineau des For\^ets, Shalabiea, Terzieva, \&\ Herbst 1998).  
Such models give two stable solutions of chemical equilibrium, 
the low ionization phase (LIP) and the high ionization phase (HIP), 
at low densities.   

[CI] line observations from objects with high visual extinction ($A_{\rm V}$) may be crucial 
to answer the question whether C$^0$ really exists deep inside molecular clouds.  
In this study, we chose DR15, an HII region in the Cygnus-X region, as a target.  
DR 15 is associated with two mid-infrared dark clouds (IRDCs; Egan et al. 1998), 
which belong to a population of compact objects with significant mid-infrared extinctions 
($A_{8 \mu{\rm m}}\!>\!2$ which corresponds to $A_{\rm V}\!\gtrsim\!200$).   
H$_2$CO observations (Carey et al. 1998) have shown that 
IRDCs are cold ($T_{\rm k}\!<\!20$ K) and dense [$n({\rm H}_2)\!>\!10^5$ cm$^{-3}$].   
Lack of mid- to far-infrared emission associated with these clouds suggests that 
they do not contain currently forming high-mass stars.   
DR15 itself is formed by one or two B0 ZAMS star(s), 
which is observed as the far-infrared source FIR 1 (Odenwald et al. 1990), 
having a total luminosity of $\sim\!3\times 10^6\, L_{\sun}$.
Assuming that all the far-infrared flux of FIR 1 was initially emitted in the ultraviolet (6--13.6 eV), 
an UV enhancement factor $G_0$ is estimated to be $\sim 550$ at 1 pc from the ionizing star(s).  

This paper presents the results of CI $^3P_1$--$^3P_0$, CO {\it J}=3--2, 
$^{13}$CO {\it J}=1--0 and C$^{18}$O {\it J}=1--0 observations of the DR 15 region.  
These data are used to study the C$^0$ column density and C$^0$/CO abundance ratio, 
and their spatial variations, especially in relation to the UV source and the IRDCs.

\section{OBSERVATIONS}

\subsection{Mt. Fuji Observations}

The CI $^3P_1$--$^3P_0$ (492.160651 GHz) and CO {\it J}=3--2 (345.795991 GHz) 
observations were performed with the Mount Fuji submillimeter-wave telescope (Sekimoto et al. 2000) 
in 1999 March and December, and 2000 January.  
The diameter of the telescope is 1.2 m, which gives 
half-power beamwidths of 2\arcmin .2 at 492 GHz and 3\arcmin .1 at 346 GHz.  
The pointing of the telescope was checked and corrected 
by observing the sun and the full moon every month, 
and its accuracy was maintained within 20\arcsec(rms) during the observing season.  
The beam efficiencies including radome loss ($\eta_{moon}$) 
were measured by observing the full moon to be 72 \%\ (492 GHz) and 75 \%\ (346 GHz), 
and the main-beam efficiencies are estimated to be 45 \%\ (492 GHz) and 67 \%\ (346 GHz).  
The telescope efficiencies and beamwidths at the observed frequencies are summarized in Table 1.  

The telescope is equipped with three SIS receivers (346 GHz/ 492 GHz/ 809 GHz) 
operated in the double-sideband (DSB) mode. 
The typical DSB system noise temperatures during the observations 
were 700 K for 492 GHz and 500 K for 346 GHz.  
Calibration of the antenna temperature was accomplished by chopping 
between an ambient temperature load and the sky.  
Reproducibility of intensities were checked by monitoring the calibration sources, 
Orion KL and DR21, and were found to be reliable 
within 9 \% ($1 \sigma$) at 492 GHz and 11 \% ($1 \sigma$) at 346 GHz \footnotemark[1] 
during the DR15 observations. 
All spectra were obtained with a 1024 channel acousto-optical spectrometer (AOS) 
which covers an instantaneous bandwidth of 700 MHz with a spectral resolution of 1.6 MHz.  
These correspond to a 430 \kms\ velocity coverage and an 1 \kms\ velocity resolution 
at the [CI] line frequency.  

\footnotetext[1]{Intensity reproducibility of Mt. Fuji submillimeter-wave telescope is limited by the effect of optical depth and variation in the temperature difference between the chopper inside the radome and the outside atmosphere.}

The [CI] observations were performed in a frequency-switching mode.  
Switching frequency was set to 80 MHz and intervals to 10 seconds.  
The on-source integration times was 100 s for each position 
and rms noise was 0.1 K in $T_A^*$. 
The CO {\it J}=3--2 observations were performed 
by position switching to a clean reference position, 
$(\alpha_{1950}, \delta_{1950}) = (20\hh 16\mm 00\ss , +43\arcdeg 00\arcmin 00\arcsec )$.  
Typical on-source integration time was 350 s which yields rms noise of 0.2 K in $T_A^*$.  
We mapped a $27\arcmin\!\times\!27\arcmin$ area in the [CI] line, 
and a $42\arcmin\!\times\!30\arcmin$ area in the CO {\it J}=3--2 line, on a 1\arcmin.5 grid.  
We subtracted baselines of the spectra by fitting fourth-order polynomial lines.

\subsection{NRO 45 m Observations}

The $^{13}$CO and C$^{18}$O {\it J}=1--0 observations have been made 
with the NRO 45 m telescope in January 2000. 
The NRO 45 m telescope has a $17\arcsec\!\pm\! 1\arcsec$ full width at half maximum beam at 110 GHz. 
Pointing errors were corrected every two hours by observing the SiO maser source UX-Cyg 
with the HEMT amplifier receiver at 43 GHz. 
The pointing accuracy of the telescope was good to $\leq\! 3\arcsec$ in both azimuth and elevation. 

We used the $2\!\times\!2$ focal-plane array SIS receiver S115Q. 
Typical system noise temperature was 400 K during the observations. 
Calibration of the antenna temperature was accomplished by a standard chopper-wheel method.  
Since the mixers in S115Q work in a quasi-single sideband mode, 
we scaled the antenna temperature for each channel referring to 
that taken with the single-beam SIS receiver S100 
equipped with a quasi-optical image rejection filter. 
Every observing day, we took spectra of DR21(CO) (Richardson et al. 1986), 
at $(\alpha_{1950}, \delta_{1950}) = (20\hh 37\mm 09.0\ss , +42\arcdeg 09\arcmin 00\arcsec )$ 
near its transit with each channel of S115Q, and calculated scaling factors 
by comparing with $T_A^{*}(^{13}{\rm CO})\!=\!12.8$ K or $T_A^{*}({\rm C^{18}O})\!=\!2.12$ K . 
We used the high-resolution acousto-optical spectrometers (AOS-Hs) 
each of which covers an instantaneous bandwidth of 40 MHz with a spectral resolution of 37 kHz. 
At the $^{13}$CO {\it J}=1--0 frequency, 
these correspond to a 110 km s$^{-1}$ velocity coverage and a 0.1 \kms\ 
velocity resolution, respectively. 

All data were obtained by position switching to a reference position, 
$(\alpha_{1950}, \delta_{1950})\!=\!(20\hh 45\mm 13\ss , +42\arcdeg 38\arcmin 52\arcsec )$. 
To conduct the survey effectively, 
three on-source positions were observed for one reference position observation. 
We mapped a $20\arcmin\times 20\arcmin$ area in the $^{13}$CO line, 
and more restricted area in the C$^{18}$O line, on a 34\arcsec\ grid.  
We also made a 17\arcsec\ grid $^{13}$CO mapping in $2\arcmin.5\times 2\arcmin.5$ areas 
around the two mid-IR dark clouds and the CO {\it J}=3--2 emission peak.  
Typically 20 seconds on source integration gave spectra with rms noise of 0.5 K ($T_A^*$) 
at a velocity resolution of 0.1 km s$^{-1}$. 
We subtracted baselines of the spectra by fitting first or third-order polynomial lines.

\section{RESULTS}

The integrated intensity map in the [CI] and CO {\it J}=3--2 line are shown 
in Fig.1 with the optical image.  
Figure 3 shows $^{13}$CO {\it J}=1--0 and C$^{18}$O {\it J}=1--0 integrated intensity maps.  
All the [CI] and CO maps are presented in units of $T_A^*$.  
In this paper, we adopt $1.0$ kpc as the distance to DR15 (Carey et al. 1998). 
At this distance, the beam sizes of the Mt. Fuji submillimeter-wave telescope 
correspond to 0.64 pc (492 GHz) and 0.90 pc (346 GHz), respectively.

\subsection{CI Distribution} 

The [CI] emission is extended over the map with patchy morphology 
being not restricted to a region adjacent to the HII region, DR15.      
It roughly follows dark patches in the optical image.    
An intense [CI] lump with a size of $8\arcmin\!\times\!6\arcmin$ ($2.3\!\times\!1.7$ pc$^2$) 
lies farther from the ionization front than the two IRDCs do.  
The [CI] lump has a 'hump' in its southern portion, near the position of G79.34+0.33,  
while G79.27+0.38 has no [CI] counterpart.  
The [CI] lump has an weak extension to the south which follows the brightest part of the optical nebula.  

\subsection{CO {\it J}=3--2 Distribution}

The CO {\it J}=3--2 emission is also extended over the map with more smooth distribution than [CI].    
It strongly peaks at the position of DR15 FIR 1 (Odenwald et al. 1990), 
which may coincides with the ionizing star(s),  
and exhibits a poor spatial correlation with the [CI] distribution.  
The displacement of the CO {\it J}=3--2 peak to that of [CI] is about 7\arcmin\ (2 pc).  
An western, weak intensity emission hump may correspond to G79.27+0.38, 
while G79.34+0.33 is included in the bulk emission associated with DR15.

\subsection{$^{13}$CO and C$^{18}$O Distributions} 

The $^{13}$CO {\it J}=1--0 and C$^{18}$O {\it J}=1--0 maps show 
two prominent cores with diameters of 2\arcmin --3\arcmin\ (0.6--0.9 pc), 
which spatially coincide with the IRDCs, G79.27+0.38 and G79.34+0.33.  
Both the $^{13}$CO and C$^{18}$O emissions closely follow dark patches in the optical image.  
The most intense CO peak lies at the G79.34+0.33 core, 
being associated with a CI counterpart.
A low intensity envelope of the $^{13}$CO emission extends over the upper half of the map, 
showing a rough spatial correlation with the [CI] lump (Fig.4).  
An weak $^{13}$CO emission peak at ($\alpha_{1950}$, $\beta_{1950}$)=(20\hh 30\mm 22\ss , +40\arcdeg 13\arcmin) 
corresponds to the northwest corner of the [CI] lump.  
No CO counterpart of the [CI] peak was found.  
Two small clumps at ($\alpha_{1950}$, $\beta_{1950}$)$\simeq$(20\hh 30\mm 40\ss , +40\arcdeg 06\arcmin) 
and another larger clump in the south correspond to 
the CO {\it J}=3--2 peak and its southern extension.   
Since the two small clumps coincide with a band of obscuration separating the optical emission nebula, 
they must be in front of the HII region.  

The two molecular cores which correspond to G79.27+0.38 and G79.34+0.33 have LTE masses of 
$2.0\times 10^3$ M$_{\sun}$ and $9.1\times 10^2$ M$_{\sun}$, respectively, 
if we assume $T_{\rm ex}\!=\!20$ K (see \S 4.2).  
These masses are similar to their virial masses, 
$1.3\times 10^3$ M$_{\sun}$ and $9.2\times 10^2$ M$_{\sun}$, 
suggesting that the cores are gravitationally bound.  
The mean densities derived from LTE masses are, $n({\rm H}_2)\!\simeq\!1.2\times 10^5$ cm$^{-3}$ 
and $\simeq\!1.3\times 10^5$ cm$^{-3}$, 
which are consistent with the densities determined from LVG analysis of H$_2$CO lines (Carey et al. 1998).

\subsection{Integrated Line Intensities} 

Figure 5 shows correlation plots between the line intensities integrated 
over the range $V_{\rm LSR}\!=\!-7.5$ \kms\ to $+7.5$ \kms\ . 
All the data are smoothed with the FWHM=2\arcmin\ Gaussian weighting function, 
and resampled to the same regular grids with 1.5\arcmin\ spacings.  
There is a rough correlation between the [CI] intensity and the $^{13}$CO {\it J}=1--0 intensity 
(correlation coefficient, $R\!=\!0.62$; Fig.4{\it a}).  
A least-squares regression to a linear relation gave a following relation; 
\begin{equation}
\int T_{\rm MB}({\rm CI}) dV = (0.66_{\pm 0.46}) + (0.25_{\pm 0.02}) \int T_{\rm MB}(^{13}{\rm CO}) dV .
\end{equation}

No correlation between the [CI] intensity and the C$^{18}$O {\it J}=1--0 intensity 
has been found ($R\!=\!0.35$; Fig.4{\it b}).  
The C$^{18}$O intensity is well correlated with the $^{13}$CO intensity ($R\!=\!0.85$; Fig.4{\it c}).  
The CO {\it J}=3--2 intensity  
shows a poor correlation with the [CI] intensity ($R\!=\!0.42$; Fig.4{\it d}).  
This seems to be remarkable, 
because the both line emissions are generally thought to arise from UV-irradiated cloud surfaces (e.g., Warin, Benayoun, \&\ Viala 1996).

\section{DISCUSSION}

\subsection{Column Densities} 

Since the critical densities of CI $^3P_1$--$^3P_0$ and CO {\it J}=1--0 lines are similar to each other 
($\simeq\,10^{3}$ cm$^{-3}$), we expect similar excitation temperatures ($T_{\rm ex}$) for the both transitions 
if C$^0$ and CO reside in the same or adjacent layers.  
We have calculated the C$^0$ and CO column densities by assuming homogeneous clouds 
in local thermodynamic equilibrium (LTE).   

The total beam-averaged atomic carbon column density is given by: 
\begin{eqnarray}
N({\rm C}^0) = 1.98\times 10^{15} \int T_{\rm MB} dV\,Q(T_{\rm ex})\,\mbox{exp}(\frac{E_1}{kT_{\rm ex}}) \nonumber \\
\left(1-\frac{J_{\nu}(T_{\rm BB})}{J_{\nu}(T_{\rm ex})}\right)^{-1}\, 
\frac{\tau_{\rm CI}}{1-e^{-\tau_{\rm CI}}}\,
\end{eqnarray}
where $Q(T_{\rm ex})$ is the ground state partition function for the carbon atom 
\begin{equation}
Q(T_{\rm ex}) = 1 + 3\,\mbox{exp}(-\frac{E_1}{kT_{\rm ex}}) + 5\,\mbox{exp}(-\frac{E_2}{kT_{\rm ex}}).      
\end{equation}
$E_1$ is the energy of the {\it J}=1 level ($E_1/k\!=\!23.6$ K), and 
$E_2$ is the energy of the {\it J}=2 level ($E_1/k\!=\!62.5$ K).   
The factor $\tau_{\rm CI}/(1-e^{-\tau_{\rm CI}})\,(\geq 1)$ is a correction factor 
for the velocity-averaged optical depth, 
and is unity at the optically thin limit ($\tau_{\rm CI}\!\ll\!1$). 
We substitute the optical depth at the peak velocity for the velocity-averaged optical depth. 
This substitution makes derived column density an upper limit, 
while the optically thin assumption gives a lower limit.   
The peak CI optical depth is given by: 
\begin{equation}
\tau_{\rm CI} = -\mbox{ln}\left[1-\frac{T_{\rm MB}}{\eta_{f} \left(J_{\nu}(T_{\rm ex})-J_{\nu}(T_{\rm BB})\right)} \right], 
\end{equation}
where $\eta_{f}$ is the beam-filling factor, and $J_{\nu}$ is the radiation temperature 
\begin{equation}
J_{\nu}(T) = \frac{h\nu/k}{\mbox{exp}(h\nu/kT) - 1}.  
\end{equation}
We assumed $\eta_{f}\!=\!1$ in the CI optical depth estimation.  
Arguments based on observations of the CI $^3P_2$--$^3P_1$/$^3P_1$--$^3P_0$ ratio 
suggest that the CI $^3P_1$--$^3P_0$ opacity is usually $\leq 1$ (Zmuidzinas et al. 1988; Genzel et al. 1988) .  
Our observations also gave $\tau_{\rm CI}< 1$,  
even for the [CI] intensity peak position ($\tau_{\rm CI}\!=\!0.6$).   
The uncertainty in the estimated C$^0$ column density 
brought by the optical depth correction may be less than 33 \% .  
Although the sub-beam filling effect may raise the optical depth,  
it does not make a significant revision to the column density, 
so far as the CI line is optically thin.   
The calibration error of submillimeter line intensity 
also brings an uncertainty in the column density (9 \%).  

The general equation for estimating the CO column density is given by: 
\begin{eqnarray}
N({\rm CO}) = \frac{5.55\times 10^{17}}{\left(\nu [{\rm GHz}]\right)^{2}}\,
\int T_{\rm MB} dV\,
T_{\rm ex}\,\mbox{exp}\left(\frac{E_{\rm u}}{k T_{\rm ex}}\right)\, \nonumber \\
\left(1-\frac{J_{\nu}(T_{\rm BB})}{J_{\nu}(T_{\rm ex})}\right)^{-1}\, 
\frac{\tau_{\rm CO}}{1-e^{-\tau_{\rm CO}}} , 
\end{eqnarray}
where $E_{\rm u}$ is the upper level energy, 
$\tau_{\rm CO}/(1-e^{-\tau_{\rm CO}})$ is a correction factor 
for the velocity-averaged optical depth of CO line ($\tau_{\rm CO}$).   
The CO optical depths are estimated by the $^{13}$CO/C$^{18}$O intensity ratios 
assuming the isotopic abundance ratio, 
[$^{13}$CO]/[C$^{18}$O]$\!=\!7.4$ (Langer \&\ Penzias 1990), 
and equal beam-filling factors.  
We used the C$^{18}$O integrated intensity to estimate the CO column density 
with the isotopic abundance [CO]/[C$^{18}$O]$\!=\!392$ (Langer \&\ Penzias 1990).  
The observed $^{13}$CO/C$^{18}$O integrated intensity ratio ($\gtrsim\!4$) implies that 
the C$^{18}$O {\it J}=1--0 transition is usually optically thin ($\tau\!\lesssim\!0.2$).  
Thus, uncertainty in the estimated CO column density 
brought by the C$^{18}$O optical depth correction is less than 12 \% .  
The excitation temperature is estimated 
from the $^{13}$CO {\it J}=1--0 peak antenna temperature 
using the relationship: 
\begin{equation}
T_{\rm MB}\left(^{13}{\rm CO}\right) = \eta_{f} \left[J_{\nu}(T_{\rm ex})-J_{\nu}(T_{\rm BB})\right]
\left(1-e^{-\tau_{13}}\right),    
\end{equation}
where $\eta_{f}$ is the beam-filling factor (assumed to be $=1$), 
$\tau_{13}$ is the $^{13}$CO optical depth at the peak velocity.   
The sub-beam filling effect may not be not significant, especially for two molecular cores, 
since their mean LTE densities are consistent with the LVG densities (\S 3.3).

Table 2 lists the C$^0$ and CO column densities toward four representative directions in DR15, 
the [CI] peak, the centers of IRDCs (G79.37+0.33, G79.27+0.38), and CO {\it J}=3--2 peak.    
The data are smoothed with the FWHM=2\arcmin\ Gaussian weighting function.  
Excitation temperatures are typically 20 K except for the CO {\it J}=3--2 peak ($T_{\rm ex}\!\simeq\!40$ K).  
The beam-averaged C$^0$/CO column density ratios toward IRDCs are $\simeq\!0.02$, 
which are much smaller than those observed in the cloud cores 
(0.16--0.32, Plume et al. 1999; Ikeda et al. 1999).  
The CO {\it J}=3--2 peak, the position coincides with DR15, 
also shows low value of $N({\rm C^0})/N({\rm CO})$ ($\simeq\!0.03$).   
In the [CI] peak, we measure $N({\rm C^0})/N({\rm CO})$ to be 0.15, 
which is similar to the typical value for the bulk of giant molecular clouds and nearby dark clouds 
(0.10--0.21, Ikeda et al. 1999; 0.08--0.25, Maezawa et al. 1999), 
although the line-of-sight extinction is far larger than those of them.

If we assume $T_{\rm ex}\!=\!18$ K and optically thin, 
the total C$^0$ number in the $^{13}$CO mapping area amounts to $1.8\!\times\!10^{55}$ atoms.  
This total C$^0$ number varies by 25 \%\ for the excitation temperature range from 15 K to 40 K.  
The optically thin assumption makes us underestimate the C$^0$ number by up to 33\% . 
The total CO number in the same area estimated under 
the optically thin assumption is $2.8\!\times\!10^{56}$ molecules. 
This is an underestimation by up to 50 \%, 
and varies by a factor of 2 for $15\!\leq\!T_{\rm ex}\!\leq\!40$ K.  
The C$^0$/CO ratio for the bulk of the DR15 cloud is $0.064\pm 0.035$.

\subsection{Origin of [CI] Emission}

\subsubsection{Does UV from DR15 Dominate C$^0$ Formation ?} 

Low C$^0$/CO ratios toward cloud cores are reflected in 
the [CI] vs. $^{13}$CO intensity correlation plot 
(Fig.5{\it a}).  
This trend has been found in the data obtained by the previous observations 
(Tauber et al. 1995; Plume et al. 1999, Ikeda et al. 1999), 
and is qualitatively consistent with the steady-state PDR models, 
that C$^0$ resides only in thin surfaces of UV-irradiated molecular clouds.    
Poor correlation of [CI] intensity with C$^{18}$O {\it J}=1--0 intensity (Fig.5{\it b}) 
also implies that [CI] emission arises from cloud surfaces and 
C$^{18}$O emission prefers cloud cores.   

However, the [CI] intensity peaks farther into the molecular cloud 
from the ionization front behind the IRDCs (Fig.1{\it b}).  
The HII region / CO core / CI lump arrangement itself is inconsistent with the expectation from steady-state, 
plane-parallel PDR models.  
In M17 and S140, similar arrangements have been reported by Keene et al. (1985).  
We have found the C$^+$/CO/C$^0$ arrangement also in the Orion KL region (Ikeda et al. 2001) 
and the $\rho$-Oph main cloud (L1688; Kamegai et al. 2001).
If the [CI] peak is shadowed by G79.37+0.33, 
the visual extinction between the ionization front and the [CI] peak position 
exceeds $80$ mag.  
The [CI] peak direction itself has a line-of-sight extinction of 20--30 mag.  
On the other hand, models for plane-parallel photodissociation regions 
have shown that the layer with $N({\rm C^0})/N({\rm CO})\!>\!0.1$ has depth smaller than 7 mag.   
Clumpy structure may help UV photons penetrate deep into molecular clouds.  
However, if we assume $n({\rm H}_2)\!=\!10^{4-5}$ cm$^{-3}$ and 
an empirical relation $N({\rm H}_2)\!=\!1.6\times 10^{21} A_{\rm V}$ cm$^{-2}$, 
$N({\rm C^0})/N({\rm CO})\!<\!0.1$ requires clumps with sizes larger than $0.4$--$4$ pc, 
which is larger than the resolution of our observations (NRO 45 m: 0.16 pc; Mt. Fuji: 0.64 pc).  

In addition, the integrated intensity of the CO {\it J}=3--2 line, 
which also arises from UV-irradiated cloud surfaces (e.g., Gierens, Stutzki, \&\ Winnewisser 1992), 
shows poor spatial correlation with the [CI] intensity (Fig.1).  
The CO {\it J}=3--2 intensity strongly peaks at DR15 FIR 1, 
having an extent of $5\arcmin\times 10\arcmin$ ($1.5\times 3$ pc$^2$), 
while the [CI] intensity peaks farther into the molecular cloud from the ionization front 
than either of the {\it J}=1--0 line of the CO isotope does.  
The situation is well illustrated in Fig.6 which shows a plot of integrated intensities 
along a line passing through the [CI] peak and DR15A\footnotemark[2] .     
A hump at DR15 in the [CI] intensity is the southern extension of the [CI] lump (\S 3.1), 
and it could correspond to the emission from thin photodissociated layer 
near surfaces of UV-irradiated clouds.  

\footnotetext[2]{A compact radio source located at the brightest part of FIR 1 (Colley 1980).}

The variation of the density could account for the observed arrangement.  
The lower density produces a lower temperature photo-dissociated layer, 
which has low rates of OH production and, consequently, 
increases C$^0$ by shutting off the CO formation route through OH 
(e.g., Hollenbach, Takahashi, \&\ Tielens 1991).  
Following this idea, the observed C$^0$ column densities require gas density around DR15 
at least an order of magnitude higher than the [CI] peak and the dense molecular core G79.37+0.33.  
G79.27+0.38 has the C$^0$ column a factor of 3 lower than that of G79.37+0.33, 
although the both have similar densities (\S 3.3) and UV environments.  
Apparently, G79.37+0.33 has a [CI] counterpart, while G79.27+0.38 does not.  
The mean H$_2$ density toward the CO 3--2 peak, 
which is estimated from the $^{13}$CO column density, 
is not so high, $n({\rm H}_2)\!\lesssim\!2\times 10^4$ cm$^{-3}$ .  
These suggest that the density is unlikely a key parameter to determine the C$^0$ column density in the DR15 region.  

The C$^0$ column density toward the CO 3--2 peak is estimated to be $9.3\!\times\!10^{16}$ cm$^{-2}$.     
According to Hollenbach, Takahashi, \&\ Tielens (1991), 
a plane-parallel PDR with $G_0\!=\!500$ and $n_0\!=\!10^4$ cm$^{-3}$ 
has a C$^0$ column density of $2.5\times 10^{17}$ cm$^{-2}$ .   
This value differs from the observed value by more than a factor of 2.  
Indeed, $G_0$ may be higher than 500 at the position close to DR15A, 
the discrepancy between theory and observation would be more significant.  
The hypothesis of unresolved, high density clumps, does not change the situation,    
since it decreases the CI column density of each clumps 
while it increases the surfaces to be consistent with the observed $^{13}$CO intensity.  
We therefore conclude that the steady-state PDR models are not able to give reasonable interpretations to 
the observed CO 3--2(HII)/CO/CI arrangement and the deficit of C$^0$ toward DR15.

\subsubsection{Origin of Atomic Carbon Unrelated to Photodissociation}

The similarity in the distributions of the [CI] and $^{13}$CO lines 
may be an indication that these lines arise from the same spatial regions.    
This implies that C$^0$ basically coexist with molecular gas, 
except for the dense regions in molecular clouds, 
such as molecular cores G79.37+0.33 and G79.27+0.38.  
This idea is supported by the results of our large-scale CI surveys of nearby molecular clouds 
(e.g., Ikeda et al. 1999).  
To account for the coexistence of C$^0$ and CO, some revisions to the steady-state PDR models or 
the application of models unrelated to photodissociation may be inevitable.  
It may be constructive to examine the other ideas for the origin of atomic carbon 
unrelated to photodissociation processes. 

Time-dependent chemistry models (e.g., Lee et al. 1996; St\"orzer, Stutzki, \&\ Sternberg 1997; Suzuki et al. 1992) 
can explain the observed [CI] and CO distributions.   
The dynamical timescale of a dense core with $n({\rm H_2})\!=\!10^5$ cm$^{-3}$ is $t_{dyn}\!\simeq\!10^5$ years, 
while its chemical timescale (timescale for conversion from C$^0$ to CO) is $t_{chem}\!\lesssim\!10^4$ years.  
Thus, the carbon in a high-density core is incorporated into CO completely before the core collapse.  
In the less dense region with $n({\rm H_2})\!=\!10^4$ cm$^{-3}$, 
the dynamical timescale and the chemical timescale become comparable [$(3\!\sim\!4)\!\times\!10^5$ years], 
and hence C$^0$ can coexist with CO for a while even after UV radiation is shielded.   
The CI-rich cloud found in Heiles Cloud 2 (Maezawa et al. 1999)   
may be the typical of such a chemically young molecular cloud. 
It is likely that high-density cores, G79.37+0.33 and G79.27+0.38, are chemically evolved, 
while less dense envelope is chemically young.  

Another idea is based on the chemical models with bistable solutions.  
These models predict that there exist two phases in the chemical equilibrium state 
at low densities ($n_{\rm H}\!\lesssim\!5500$ cm$^{-3}$; Pineau des For\^ets, Roueff, \&\ Flower 1992).   
The two stable solutions referred to as the high ionization phase (HIP) 
and the low ionization phase (LIP) lead into two different C$^0$/CO ratios 
in clouds with identical density and temperature.  
This means that the gas-phase abundance of C$^0$ can be large (C$^0$/CO=0.1--0.2) in HIP, 
even in the absence of a UV field.   
For the cloud cores, where the density exceeds the critical value for bistability, 
the system necessarily take a solution with a low C$^0$/CO ratio ($\leq 0.01$).  
The chemical bistability can be applicable to the DR15 case, 
if we consider the [CI] peak as HIP and the cores of G79.37+0.33 and G79.27+0.38 as LIP.  
The bistability model can also account for the ubiquitous value of the C$^0$/CO ratio (0.1--0.2) 
over the bulk of molecular clouds, although its applicability is still controversial.  

\section{CONCLUSIONS}

We have studied the distributions of the CI $^3P_1$--$^3P_0$ line, 
the CO {\it J}=3--2 line, as well as of the {\it J}=1--0 transition of $^{13}$CO and C$^{18}$O 
toward DR15, an HII region associated with two IRDCs. 
Neutral atomic carbon is extended over the molecular clouds traced by the $^{13}$CO {\it J}=1--0 line.  
The two dense molecular cores with masses of $2.0\times 10^3$ M$_{\sun}$ and $9.1\times 10^2$ M$_{\sun}$, 
which correspond to the IRDCs, have been found in the $^{13}$CO and C$^{18}$O data.  
The [CI] integrated intensity peaks farther into the molecular cloud 
from the ionization front behind the IRDCs.  
The column density ratio $N({\rm C^0})/N({\rm CO})$ toward the [CI] peak is 0.15,  
which is similar to those observed in giant molecular clouds and nearby dark clouds, 
while the dense cores (IRDCs) exhibit very low ratios ($\sim\!0.02$).  
The CO {\it J}=3--2 emission strongly peaks at the DR15 FIR 1, 
exhibiting a poor spatial correlation with the [CI] distribution.  
The CO {\it J}=3--2(HII)/CO/CI arrangement and the deficit of C$^0$ toward DR15  
are inconsistent with the prediction of PDR models, 
and suggest an origin for interstellar atomic carbon unrelated to photodissociation.  
Models with time-dependent chemistry or chemical bistability could account for 
the observed H$^+$/CO/C$^0$ arrangement and the C$^0$/CO ratios as well.

\acknowledgments
We are grateful to the NRO staff for excellent support in the 45 m observations. 
This study is supported by Grant-in-Aid from the Ministry of Education, Science, 
and Culture (07CE2002, 11304010, and 12740119)

\clearpage

\figcaption[f1.eps]{
({\it a}) An optical image of DR15 from the Digitized Sky Survey (DSS).  
The data are taken from the SkyView web page.  
A white triangle denotes the position of the brightest far-infrared source (DR15 FIR 1).  
White circles represent the positions of catalogued infrared dark clouds, 
G79.27+0.38 and G79.34+0.33 .
({\it b}) A map of the CI $^3P_1$--$^3P_0$ emission integrated over the velocity range 
$\VLSR= -7.5$ to $+7.5$ \kms, smoothed with the FWHM=120\arcsec\ Gaussian weighting function.  
Contours are set at intervals of 0.5 K \kms .  
({\it c}) A map of CO {\it J}=3--2 emission integrated over the velocity range 
$\VLSR= -7.5$ to $+7.5$ \kms, smoothed with the FWHM=120\arcsec\ Gaussian weighting function.  
Contours are set at intervals of 1.5 K \kms .  
\label{fig1}}

\figcaption[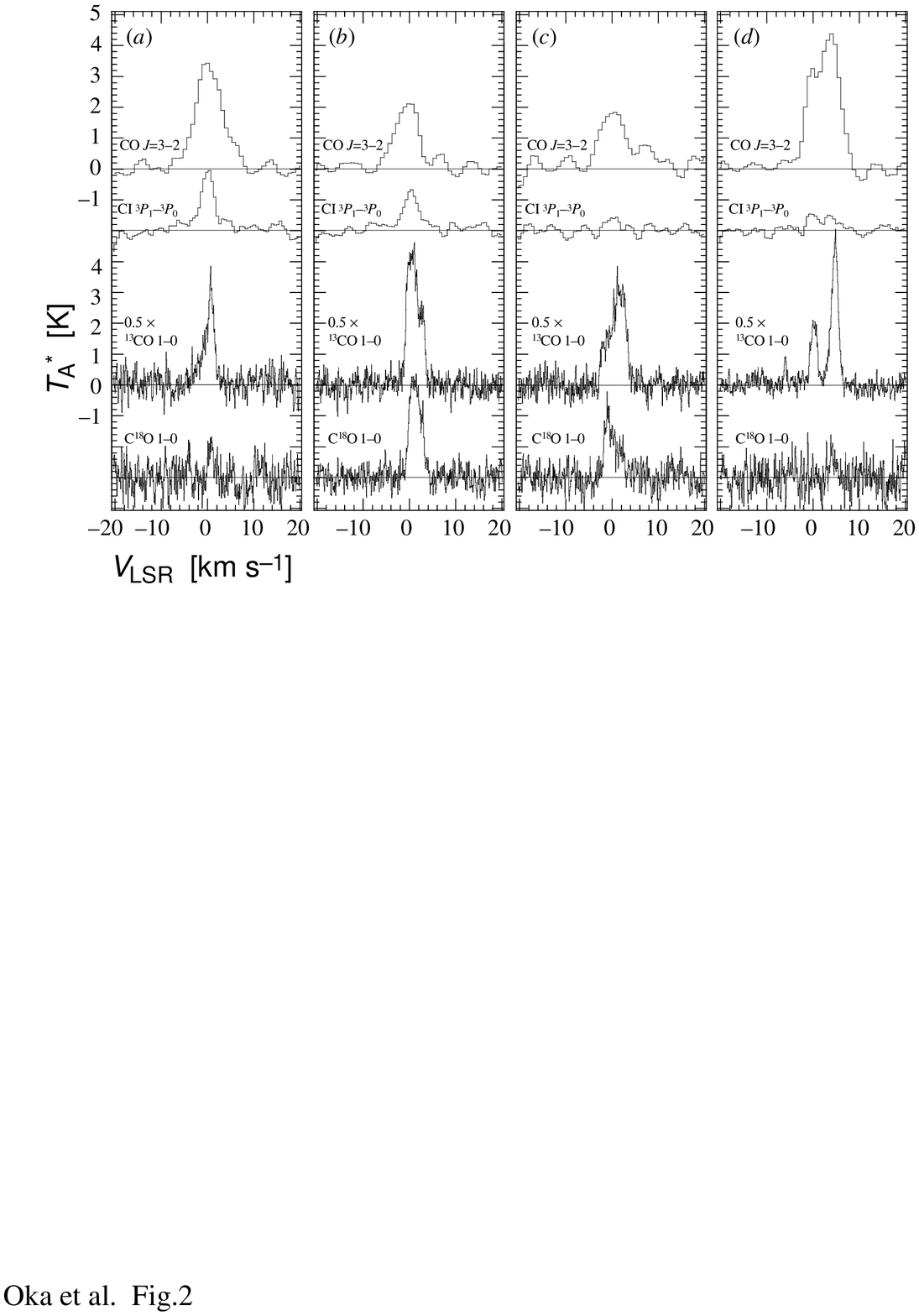]{
Observed spectra of CI $^3P_1$--$^3P_0$, CO {\it J}=3--2, 
$^{13}$CO {\it J}=1--0,  C$^{18}$O {\it J}=1--0 lines toward 
the CI peak ({\it a}), 
the centers of G79.37+0.33 ({\it b}), 
G79.27+0.38 ({\it c}), 
and the CO {\it J}=3--2 peak ({\it d}).  
\label{fig2}}

\figcaption[f3.eps]{
({\it a}) A map of the $^{13}$CO {\it J}=1--0 emission integrated over the velocity range 
$\VLSR= -7.5$ to $+7.5$ \kms, smoothed with the FWHM=45\arcsec\ Gaussian weighting function.  
Contours are set at intervals of 2 K \kms .  White contours begin at 26 K \kms .
({\it b}) A map of the C$^{18}$O {\it J}=1--0 emission.  
Velocity range and smoothing width are the same as panel {\it a}).   
Contours are set at intervals of 0.8 K \kms .  White contours begin at 7.2 K \kms .
({\it c}) Map of the $^{13}$CO {\it J}=1--0 integrated intensity overlaid on the DSS image.  
Contours are set at intervals of 4 K \kms .  
({\it d}) Map of the C$^{18}$O {\it J}=1--0 integrated intensity overlaid on the DSS image.  
Contours are set at intervals of 1.6 K \kms . 
\label{fig3}}

\figcaption[f4.eps]{
({\it a}) Map of the CI $^3P_1$--$^3P_0$ integrated intensity (contour) 
overlaid on the map of the $^{13}$CO {\it J}=1--0 integrated intensity (gray scale).  
({\it b}) Map of the [CI] integrated intensity (contour) 
overlaid on the map of the C$^{18}$O {\it J}=1--0 integrated intensity (gray scale).  
\label{fig4}}

\figcaption[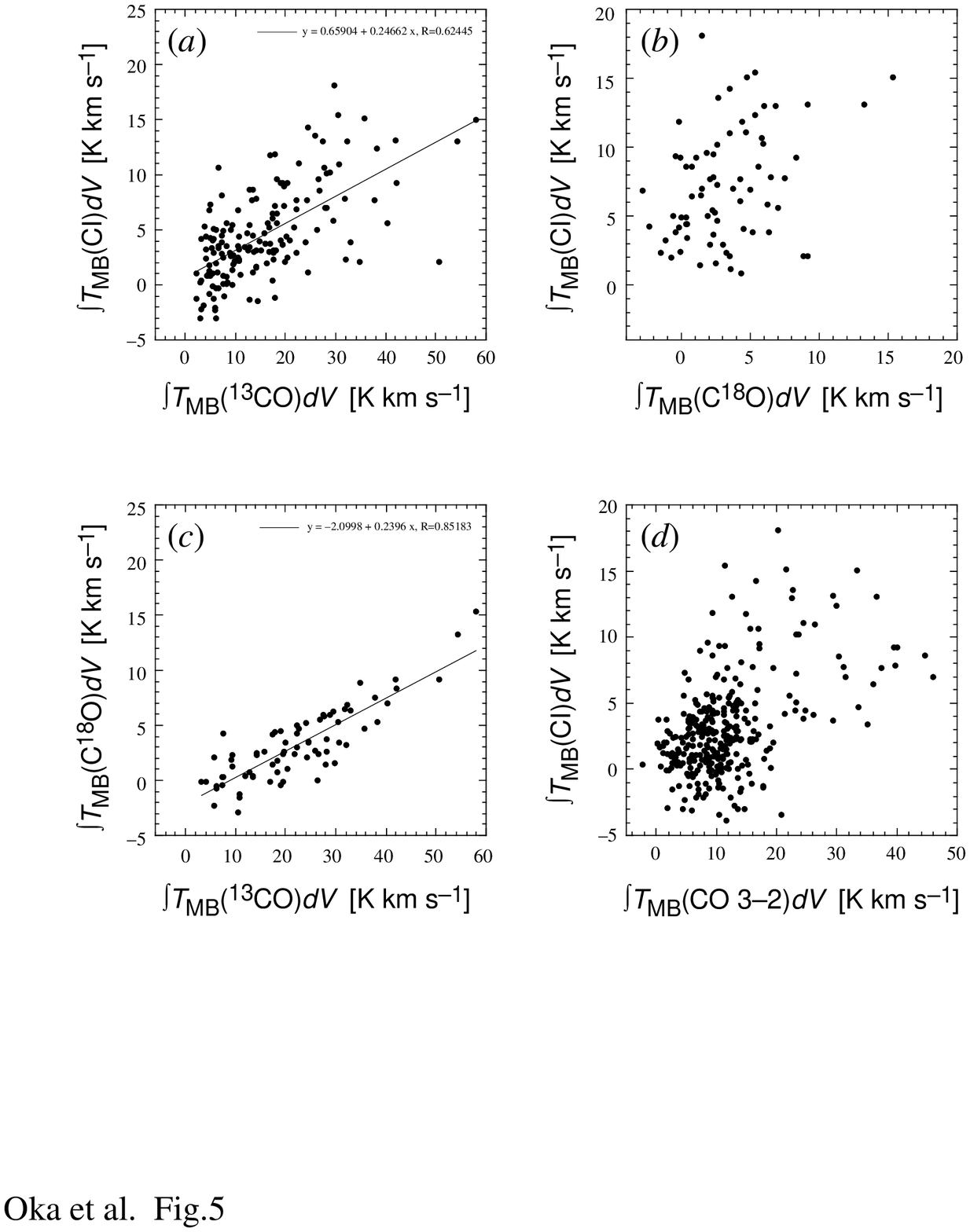]{
({\it a}) Plot of the [CI] integrated intensity vs. the $^{13}$CO {\it J}=1--0 integrated intensity.  
({\it b}) Plot of the [CI] integrated intensity vs. the C$^{18}$O {\it J}=1--0 integrated intensity.  
({\it c}) Plot of the C$^{18}$O {\it J}=1--0 integrated intensity vs. the $^{13}$CO {\it J}=1--O {\it J}=1--0 integrated intensity.  
({\it d}) Plot of the [CI] integrated intensity vs. the CO {\it J}=3--2 integrated intensity.  
\label{fig5}}

\figcaption[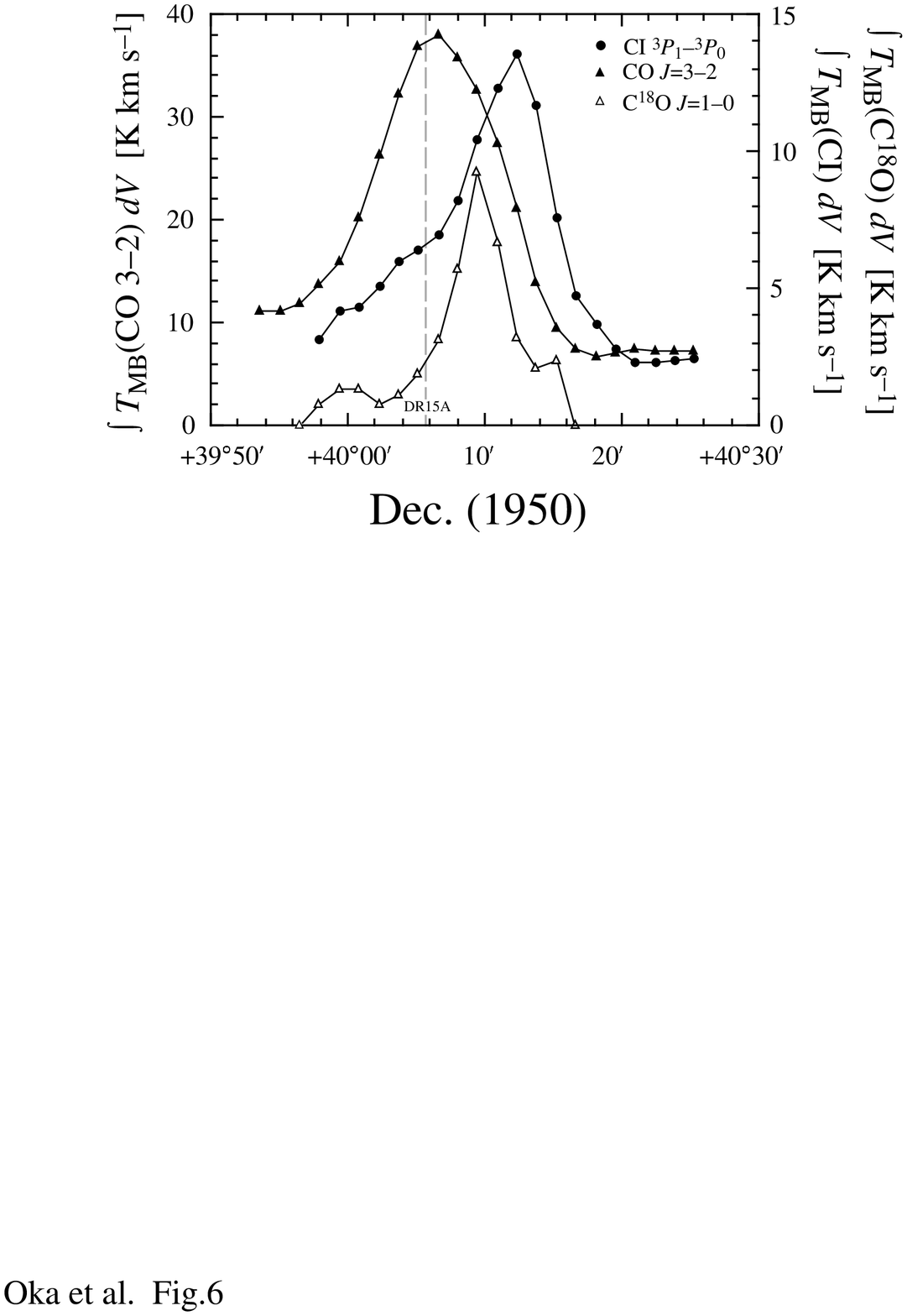]{
Comparison of the [CI] (filled circles), CO {\it J}=3--2 (filled triangles), 
and C$^{18}$O {\it J}=1--0 (open triangles) integrated intensities.  
The cross section passes DR15A (Colley 1980), 
$(\alpha_{1950}, \delta_{1950})\!=\!(20\hh 30\mm 40.22\ss , +40\arcdeg 05\arcmin 47.3\arcsec )$,  
and the [CI] peak position, 
$(\alpha_{1950}, \delta_{1950})\!=\!(20\hh 30\mm 36.0\ss , +42\arcdeg 12\arcmin 37.3\arcsec )$.  
\label{fig6}}

\clearpage

\begin{deluxetable}{lcccccccc}
\footnotesize
\tablecaption{Telescope parameters \label{tbl-1}}
\tablewidth{0pt}
\tablehead{
\colhead{Line} & \colhead{Transition} & \colhead{Freq. [GHz]} & 
\colhead{Telescope} &\colhead{HPBW} &
\colhead{$T_{\rm sys}$ [K]} & \colhead{$\eta_{\rm fss}$ ($\eta_{\rm moon}$)}  & 
\colhead{$\eta_{\rm mb}$} & \colhead{Spectra}
}
\startdata
CI & $^3P_1$--$^3P_0$ & 492.160651 & Mt. Fuji & 2\arcmin.2 & 900 & 0.72 & 0.45\tablenotemark{\dag} & 539 \\
CO & {\it J}=3--2 & 345.795989 & Mt. Fuji & 3\arcmin.1 & 500 & 0.75 & 0.67\tablenotemark{\dag} & 324 \\
$^{13}$CO & {\it J}=1--0 & 110.201353 & NRO 45 m & 17\arcsec & 400 & 0.58 & 0.48 & 1088 \\
C$^{18}$O & {\it J}=1--0 & 109.782182 & NRO 45 m & 17\arcsec & 400 & 0.58 & 0.48 & 624 \\
\tablenotetext{\dag}{The effective main-beam efficiencies including the effect of optical depth and the average temperature difference between the chopper inside the radome and the outside atmosphere.}
\enddata
\end{deluxetable}

\begin{deluxetable}{lccccc}
\footnotesize
\tablecaption{Column Densities \label{tbl-2}}
\tablewidth{0pt}
\tablehead{
\colhead{} & \colhead{$T_{\rm ex}$ [K]} & \colhead{$N({\rm CO})$ [cm$^{-2}$]} & 
\colhead{$N({\rm C}^0)$ [cm$^{-2}$]} &\colhead{$A_{\rm V}$\tablenotemark{\dag}} 
& \colhead{$N({\rm CI})/N({\rm CO})$} 
}
\startdata
CI peak\tablenotemark{a} & 18.0 & $2.0\!\times\!10^{18}$ & $2.8\!\times\!10^{17}$ & 21--30 & 0.15 \\
G79.37+0.33\tablenotemark{b} & 22.8 & $8.4\!\times\!10^{18}$ & $1.9\!\times\!10^{17}$ & 87--130 & 0.022 \\
G79.27+0.38\tablenotemark{c} & 18.1 & $4.1\!\times\!10^{18}$ & $7.2\!\times\!10^{16}$ & 44--64 & 0.017 \\
CO 3--2 peak\tablenotemark{d} & 37.8 & $3.3\!\times\!10^{18}$ & $9.3\!\times\!10^{16}$ & 51 & 0.028 \\
\tablenotetext{a}{$(\alpha_{1950}, \delta_{1950}) = (20\hh 30\mm 36.0\ss , +40\arcdeg 12\arcmin 37.3\arcsec )$.}
\tablenotetext{b}{$(\alpha_{1950}, \delta_{1950}) = (20\hh 30\mm 33.1\ss , +40\arcdeg 09\arcmin 20.3\arcsec )$.}
\tablenotetext{c}{$(\alpha_{1950}, \delta_{1950}) = (20\hh 30\mm 07.9\ss , +40\arcdeg 08\arcmin 12.3\arcsec )$.}
\tablenotetext{d}{$(\alpha_{1950}, \delta_{1950}) = (20\hh 30\mm 43.9\ss , +40\arcdeg 06\arcmin 37.3\arcsec )$.}
\tablenotetext{\dag}{Estimated from the CO column density using empirical relations, $N({\rm C^{18}O})\!=\!2.5\times 10^{14} (A_{\rm V}-1.5)$ (Cernicharo \&\ Gu\'elin 1987) and $N({\rm H})\!=\!1.6\times 10^{21} A_{\rm V}$.}
\enddata
\end{deluxetable}

\end{document}